\documentclass{article}
\usepackage{spconf,amsmath,graphicx}
\usepackage{dsfont}
\usepackage{booktabs}
\usepackage{multirow}
\usepackage{float}
\usepackage{cite}
\usepackage{amssymb}
\usepackage{comment}
\usepackage{hyperref}

\DeclareMathOperator*{\argmin}{arg\,min}
\title{Improving the transferability of speech separation by meta-learning}
\name{ Kuan-Po Huang$^1$$^\star$ \qquad Yuan-Kuei Wu$^2$$^\star$  \qquad Hung-yi Lee$^3$
\thanks{$^\star$The two first authors made equal contributions.}}
\address{$^{1}$Graduate Institute of Computer Science and Information Engineering, National Taiwan University\\
$^{23}$Graduate Institute of Communication Engineering, National Taiwan University\\
$^{123}$\{r09922005, f07942100, hungyilee\}@ntu.edu.tw}

\begin{document}
\ninept
\maketitle

\begin{abstract}
Speech separation aims to separate multiple speech sources from a speech mixture. Although speech separation is well-solved on some existing English speech separation benchmarks, it is worthy of more investigation on the generalizability of speech separation models on the accents or languages unseen during training. This paper adopts meta-learning based methods to improve the transferability of speech separation models. With the meta-learning based methods, we discovered that only using speech data with one accent, the native English accent, as our training data, the models still can be adapted to new unseen accents on the Speech Accent Archive. We compared the results with a human-rated native-likeness of accents, showing that the transferability of MAML methods has less relation to the similarity of data between the training and testing phase compared to the typical transfer learning methods. Furthermore, we found that models can deal with different language data from the CommonVoice corpus during the testing phase. Most of all, the MAML methods outperform typical transfer learning methods when it comes to new accents, new speakers, new languages, and noisy environments.
\end{abstract}
\section{Introduction}
\label{sec:intro}
Speech applications such as automatic speech recognition (ASR), speaker verification and spoken language understanding bring convenience to our daily life. Nonetheless, while multiple speakers talk simultaneously, the models fail to work appropriately on these tasks. Thus, a speech separation (SS) model, which can separate sources from the mixtures, is crucial to this situation. Recent breakthroughs in speech separation\cite{luo2019conv, luo2020dual, chen2020dual, tzinis2020sudo, zeghidour2020wavesplit, chen2021continuous, subakan2021attention, nachmani2020voice} have resulted in high performances on the English benchmark corpus.

We believe that applying a separation model to various accents and languages is essential in real-world applications. However, the need of a large amount of training data limits the applicability of the speech separation model. Speech data with a particular accent or language may be hard to collect. These training data is either prohibitively expensive or unavailable for most languages and accents, causing the low resource problem in the speech separation task.

Model training across different accents or even languages has been explored in speech applications. 
For acoustic modeling, multi-task learning\cite{tong2017investigation,ghorbani2019leveraging,jain2018improved,rao2017multi,zhu2020multi,zhang2021e2e} is applied to improve speech recognition for accented speech.
\cite{Sim2018} adopted selective fine-tuning to alleviate the domain-mismatch problem.
\cite{li2018multi} fed the dialect-specific information into the model to help perform multi-dialect speech recognition, where dialects are defined as variations of the same language.
\cite{huang2020cross} pretrained the model with English data and adapted the model to different accents or languages by transfer learning. 
Meta-learning, also known as learning to learn, has been widely used to solve low-resource problems in speech processing tasks.
By utilizing the meta-learning algorithm, \cite{winata2020learning} trained a speech recognition model with multiple accents and tested it on unseen accents, while \cite{hsu2020meta} trained a multilingual speech recognition model and tested on unseen languages. 

Meta-learning has been used in speech separation. 
\cite{wu2021one} is the first paper using MAML\cite{finn2017model} (Model-Agnostic Meta-Learning) and its simplification ANIL\cite{raghu2019rapid} (Almost No Inner Loop) algorithm to adapt the separation model to new speakers. \cite{huang2021multiaccent} trained their model with multiple accents and adapted their model to unseen accents. 
Different from \cite{wu2021one} and \cite{huang2021multiaccent}, in this paper, we focused on adapting to different accents and languages while only one accent and language is available during the training stage. 
Moreover, we measured whether the accent similarities of the training set and testing set have relation with the performance. 
We also extended the mixtures of speech to a 3-speaker scenario. 
We made several contributions,
\begin{itemize}
    \item To our best knowledge, we are the first to train a speech separation model with only native English speakers and transfer the model to other different accents and languages of speech with meta-learning.
    \item We found out that the transferability of speech separation models has little relation to the accent similarities between the source data and target data, and MAML methods can even lower this relation.
\end{itemize}

\section{Methods}
\label{sec:methods}

\subsection{Meta-learning}
In meta-learning, there is a set of source tasks, denoted as $\mathcal{T}_{\text{source}} = \{T^s_1, T^s_2, \dots, T^s_n\}$ and a set of target tasks, denoted as  $\mathcal{T}_\text{target} = \{T^t_1, T^t_2, \dots, T^t_m\}$. 
The main concept of meta-learning is to learn the hyper-parameters, for example, the initialized parameters, from the source tasks  $\mathcal{T}_{\text{source}}$, in the hope that the model has good ability to quickly adapt to the target tasks $\mathcal{T}_\text{target}$. 
For each task $T_j$, we split the data into two sets, the support set $T^{\text{sup}}_j$ and query set $T^{\text{que}}_j$. 
In the speech separation scenario, the support set and query set consist of some speech mixtures of different speakers. 
This paper focuses on the one-shot learning setting; that is, only one mixture is used in the support set.

In this paper, the source tasks are constructed only with utterances from standard English speakers. 
Assume that there are $n$ standard English speakers $\mathcal{P} = \{p_1, p_2, \dots, p_n\}$, and $\mathcal{S}^i = \{\mathbf{s}^i_j\}$ denotes speech signals from speaker $p_i$. 
We first select a subset of speakers $\mathcal{Q} \subseteq \mathcal{P}$. 
By mixing the speech signals between different speakers in $\mathcal{Q}$ with various SNR (signal-to-noise ratio) levels, we obtain a source task $T^s_i$. 

The target tasks are formed by utterances in different English accents or other languages. 
The accents and languages are assumed unavailable for training when constructing the target tasks; that is, they are not involved in finding the hyper-parameters for quick adaptation on the source tasks.
For the source tasks, we constructed them under the scenario as below. 
A task consists of utterances with the same accent. 
Assume there are $m$ accents $\mathcal{A} = \{a_1, a_2, \dots, a_m\}$, and $\mathcal{V}^i = \{\mathbf{v}^i_j\}$ denotes speech signals of accent $i$. 
We selected an accent $a_i \in \mathcal{A}$, and mixed the utterances in $\mathcal{V}^i$ with the same method as constructing the source tasks. 

Section \ref{subsec:meta_task_cons} elaborates more details about meta task construction.

\subsection{MAML}
The meta-learning approach, MAML\cite{finn2017model}, is used in our work. 
MAML finds a set of well-initialized parameters $\theta$ for the model from the source tasks, which is easy to be adapted in the target task. 
The overall process can be divided into two phases, the meta-learning phase and the fine-tuning phase.

\subsubsection{Meta-learning phase}
The algorithm can be divided into two optimization loops, namely inner loop and outer loop. 
 We draw a batch of tasks $\{ T_1, \dots , T_B\}$ with batch size $B$ from the  source task set $\mathcal{T}_{\text{source}}$. 
 In the inner loop, we first conduct the task-specific learning process and perform the inner loop separately for each task. For each task $T_j$, the learning process can be formulated as finding the parameters that minimize the Si-SNR loss with uPIT\cite{kolbaek2017multitalker} over the support set $T^{\text{sup}}_j$ in the task $T_j$,
\begin{equation}
    \theta_j = \arg\min_{\theta} \mathcal{L}_{T^{\text{sup}}_j}(\theta)
\end{equation}
We calculate $\theta_j$ by using one or multiple gradient descent steps. For simplicity of notation, we formulated it in one gradient descent step with meta-learning rate $\alpha$ as below.
\begin{equation}
\label{eq:adaptation}
    \theta_j \leftarrow \theta - \alpha \nabla_{\theta}\mathcal{L}_{T^{\text{sup}}_j}(\theta)
\end{equation}

In the outer loop, we utilize the task-specific models, which are computed by the inner loop, to update the initialized parameters for the generalized model. We define the meta loss function over the query set $T^{\text{que}}_j$ as,
\begin{equation}
    \mathcal{L}_{\text{meta}}(\theta) = \sum_{j=1}^{B} \mathcal{L}_{T^{\text{que}}_j}(\theta_j)
\end{equation}
The ultimate goal is to find the generalized model, which has the ability to quickly adapt to each task. This can be formulated using the meta loss function. 
\begin{equation}
    \theta^\star = \argmin_{\theta} \mathcal{L}_{\text{meta}}(\theta)
\end{equation}
Finally, we perform gradient descent to update the model parameters.
\begin{equation}
    \label{eq:meta-update}
    \theta \leftarrow \theta - \beta \nabla_{\theta}\sum_{j=1}^{B} \mathcal{L}_{T^{\text{que}}_j}(\theta_j)
\end{equation}

\subsubsection{Fine-tuning phase}
During the fine-tuning phase, we intend to adapt the model to the target tasks. 
In MAML, the adaptation procedure in the fine-tuning phase is the same as the inner loop in the meta-learning phase.
In the one-shot learning setup, only one data sample in the support set of a target task $T_i$ is available for adaptation. 
We obtain the target task-specific model $\theta_i$ by performing gradient descent on the well-initialized model $\theta$ (obtained from the meta-learning phase). 
The overall procedure is:
\begin{equation}
    \label{eq:finetune}
    \theta_i \leftarrow \theta - \alpha \nabla_{\theta} \mathcal{L}_{T^{\text{sup}}_i}(\theta)
\end{equation}

\section{Experiment Setups}
\label{sec:exp}

\subsection{Meta task construction}
\label{subsec:meta_task_cons}
We followed the task construction method in the previous work \cite{wu2021one}. For a 2-speaker setting, every 2 different speakers with the same accent can form a meta task. 
This meta task is constructed by sampling three source utterances from each speaker and mixing the source utterances mutually. This results in $3\times 3 = 9$ mixtures. Since our paper performs one-shot learning, we only sample one of the  9 mixtures for the support set. 
Notice that same source utterances should only appear in the support or query set but not both. Consequently, the 4 mixtures without containing source utterances in the support set form the query set. 
The same idea is applied to the 3-speaker separation setting. 
3 different speakers can form a meta task under a 3-speaker setting. 
In a 3-speaker meta task, there will be one mixture in the support set and $2\times 2\times 2 = 8$ mixtures in the query set.

\begin{table*}[t]
\centering
\setlength\tabcolsep{4pt}
\renewcommand{\arraystretch}{0.8}
\begin{tabular}{ccccccccccccccc|c}
\toprule
model & method & f.t. & noise & Al  & Ga & Ha & It & Ku & Li & Me & Ru & Ta & Th  & overall & corr \\
\midrule
conv  & joint  & - & - & 7.96          & 10.55         & 5.94            & 9.03             & 9.10           & 6.26       & 7.98     & 9.53    & 10.45 & 9.85 & 8.86  & 0.11 \\
conv  & joint  & \checkmark & - & 8.47          & 9.79          & 6.32            & 9.10              & 9.09          & 6.5        & 9.86     & 9.27    & 10.46 & 9.85  & 8.91 & -0.37\\
dprnn & joint  & - & - & 9.80           & 12.10          & 8.95            & 10.56            & 10.46         & 7.14       & 10.27    & 10.43   & 11.54 & 11.16 & 10.36  &  0.23\\
dprnn & joint  & \checkmark & - & 9.93          & 11.85         & \textbf{9.12}   & 10.63            & 10.19         & 7.58       & 9.59      & 10.46   & 11.53 & 11.13  & 10.37 & -0.37\\
\midrule
conv  & FOMAML & \checkmark & - & 9.22          & \textbf{11.49}& 6.62            & 10.40    & 9.88          & \textbf{8.62} & 9.49       & 10.46   & \textbf{11.82} & 10.72 & 9.94   & 0.01\\
conv  & MAML   & \checkmark & -  &  \textbf{9.60}& 11.20          & \textbf{8.65}   & \textbf{10.62}            & \textbf{10.56} & 8.48         & \textbf{11.51}  & \textbf{10.73}    & 11.32  &\textbf{10.81}  & \textbf{10.34} & 0.05\\
dprnn & FOMAML & \checkmark & - & \textbf{10.32}& \textbf{13.00}   & 7.45            & \textbf{11.53}   & \textbf{10.55}& \textbf{8.66} & \textbf{12.27}     & \textbf{12.20}   & \textbf{13.11} & \textbf{11.97} &  \textbf{11.10} & -0.08\\
\midrule
\midrule
conv  & joint  & - & \checkmark & 6.33 & 7.9	& 4.33	& 7.16	& 7.08	& 4.69	& 6.38	& 7.42	& 8.31	& 7.92 & 6.94 & 0.12\\
conv  & joint  & \checkmark & \checkmark & 6.8 & 6.92 & 5.1 & 7.23 & 7.26 & 5.41 & 8.07 & 7.74 & 8.29 & 7.94 & 7.16 & 0.15\\
dprnn & joint  & - & \checkmark & 7.52 & \textbf{9.54}	& 6.78 & 8.27 & 8.23 & 5.39	& 8.65	& 8.12	& 8.90	& 8.74 & 8.07 & 0.30\\
dprnn & joint  & \checkmark & \checkmark & 7.93	& 9.17	& 6.6 & 8.14 & 8.06 & 5.89 & 6.68	& 8.19 & 8.66 & \textbf{8.86} & 8.06 & 0.28\\
\midrule
conv  & FOMAML & \checkmark & \checkmark & \textbf{7.77}	& 8.79	& 5.17	& 8.47	& 8.29	& \textbf{7.22}	& \textbf{9.14}	& 8.59	& 8.79	& 8.66	& 8.09 & 0.07\\
conv  & MAML   & \checkmark & \checkmark & 7.76 & \textbf{9.91} & \textbf{5.40} & \textbf{9.03} & \textbf{8.88}	& 7.18 & 8.57 & \textbf{9.58} & \textbf{10.22} & \textbf{9.48} & \textbf{8.71} & -0.04\\
dprnn & FOMAML & \checkmark & \checkmark & \textbf{8.03}	& 9.00	& \textbf{7.03}	& \textbf{8.53}	& \textbf{8.73}	& \textbf{6.83}	& \textbf{9.09}	& \textbf{8.68}	& \textbf{8.96} & 8.84	& \textbf{8.41} & 0.07 \\
\bottomrule
\end{tabular}

\caption{Results of joint training and MAML methods under the 2-speaker setting on the 10 target accents of the Speech Accent Archive. For the Conv-TasNet(conv) and Dual-path RNN(dprnn) model, we mark the highest Si-SNRi values bold. The right most column is the Pearson correlation between accent similarities and performance.
}
\label{tab:accent_result_2spk}
\end{table*}

\subsection{Dataset}
In this paper, we used two speech datasets, the Speech Accent Archive\cite{Wei2014speech} and the CommonVoice corpus\cite{ardila2019common}.
The Speech Accent Archive contains more than 200 kinds of accents and currently has more than 2900 utterances. Each utterance is spoken in English by different speakers. We selected accents that include more than 3 utterances of different speakers since this is the minimal requirement to generate a 3-speaker mixture. 
The training tasks consist of only English-accent speech. 
We constructed approximately 2200 tasks for the training set. 
For the testing set, we selected another 10 accents of speech and constructed about 400 tasks. The ten accents used for testing are Albanian(Al), Ga, Hausa(Ha), Italian(It), Kurdish(Ku), Lithuanian(Li), Mende(Me), Russian(Ru), Tamil(Ta), and Thai(Th). The rest of the accents are used for the validation set.

The other speech dataset involved is CommonVoice. CommonVoice is a huge multilingual speech dataset. It currently has 60 languages, each with several utterances. We used the testing split of the three kinds of Chinese speech, zh-CN(China), zh-HK(Hong Kong), and zh-TW(Taiwan) as our target data. We constructed about 400 tasks for each of the three. We adopted this dataset because we wanted to test our proposed method on speech with a different language other than English. 
Since the three sets are all Chinese speech of different regions, we view them as different accents as a matter of course.

To test the experiments under a noisy setting, we added Musan\cite{snyder2015musan} noise to the test sets of the two aforementioned datasets. \footnote{Datasets available at \href{https://github.com/nobel861017/MTSS}{https://github.com/nobel861017/MTSS}.}

\subsection{Models}
The models we used in this paper are Conv-TasNet\cite{luo2019conv} and Dual-path RNN\cite{luo2020dual}. Both of them are mask-based speech separation models. For both of the models, we used the model configuration with the best performance reported in the original papers.

\subsection{MAML and FOMAML}
In the experiments, we used MAML to perform transfer learning. We also trained with FOMAML\cite{finn2017model}, a simplified version of gradient calculation. FOMAML saved a lot of time compared to MAML since it does not calculate second-order derivatives. 
For the Conv-TasNet model, we trained it with MAML and FOMAML. However, for the Dual-path RNN model, we only trained it with FOMAML since training with MAML takes too much computational time. Moreover, we found that the results of training with FOMAML are high enough. Thus, we did not do further experiments on the Dual-path RNN model with MAML.

\subsection{Accent similarity and performance}
\label{subsec:acccent_sim}
In this experiment, we want to know whether the similarities of the accents between the training set and testing set affect the performance of transfer learning on the speech separation task. 
We adopted the human accent ratings mentioned in \cite{bartelds2020neural}. 
This rating is constructed by taking 50 speech samples from the Speech Accent Archive and conducting a survey on how likely the speech samples sound like spoken by native English speakers. Multiple participants gave native-likeness ratings ranging from 1(very foreign sounding) to 7(native American English sounding). Since the training set only contains native English speakers, this native-likeness ratings can represent the similarity of accents in the testing set with respect to the training set. 
We calculated the Pearson correlation between the native-likeness ratings and the Si-SNRi values of the separated waveforms for each model and training method.

\subsection{Baseline: Joint training} 
In the setting of this paper, the target domain has speech of accents or languages different from the source domain, while the source domain has only standard English speakers.
The scenario can also be solved by typical transfer learning.
In transfer learning, a model is learned from the source domain data and adapted by limited examples from the target domain.
Learning from the source domain data is equivalent to merging all the data in the meta training tasks together to form a training set and optimize the model on it. 
Hence, the typical transfer learning is called \textit{joint training} in this paper.
In our setting here, only a training example from the target domain is available for adaptation.
To adapt the model to the target domain, we tested the jointly trained models with two methods. The first method is directly testing the model on the target data without fine-tuning. The second method is to adapt the model with the support set in a task and test its performance on the query set. 
More specific details are elaborated in Section \ref{sec:gen_exp_set}.

\subsection{General experiment settings}
\label{sec:gen_exp_set}
For every model, we trained for 200 epochs with the Adam optimizer with 0.001 learning rate and halved the learning rate whenever there is no improvement for three consecutive steps on the validation set. For the training methods MAML and FOMAML, we set the meta batch size as 3 and the fast adapt learning rate as $\alpha = 0.01$. For joint training, we adapted the trained models with a learning rate $\beta$ within a range from $1\mathrm{e}{-6}$ to $1\mathrm{e}{-1}$ and reported the best performance.

\section{Results}
\label{sec:results}

\begin{table*}[t]
\centering
\setlength\tabcolsep{4pt}
\renewcommand{\arraystretch}{0.8}
\begin{tabular}{ccccccccccccccc|c}
\toprule
model & method & f.t. & noise & Al   & Ga    & Ha   & It    & Ku   & Li   & Me    & Ru   & Ta    & Th    & overall & corr \\
\midrule
conv  & joint  & - & -  & 5.47 & 8.67  & 2.97 & 6.93  & 6.39 & 3.4  & 4.71  & 6.47 & 8.45  & 8.04  & 6.16  & 0.18  \\
conv   & joint  & \checkmark & -   & 5.59 & 7.89  & 3.24 & 6.98  & 5.96 & 3.21 & 4.45  & 6.55 & 7.67  & 7.82  & 6.05  &  -0.38\\
dprnn & joint  & -  & -   & 6.13 & 9.84  & 4.49 & 8.74  & 7.24 & 5.53 & 7.91  & 7.99 & 9.76  & 10.06 & 7.58  &-0.13  \\
dprnn & joint  & \checkmark & -    & 6.31 & 9.54  & 4.29 & 8.62  & 7.36 & 5.5  & 9.05  & 8.01 & 9.68  & 10.10  & 7.57  & -0.5 \\
\midrule
conv  & FOMAML & \checkmark & -    & 6.61 & 8.6   & 4.43 & 7.88  & 7.2  & 5.28 & 7.65  & 7.44 & 8.31  & 8.31  & 7.04  & -0.04 \\
conv  & MAML & \checkmark & -      & \textbf{8.05} &\textbf{9.93}  & \textbf{5.72}  & \textbf{9.57}  & \textbf{8.92}  & \textbf{6.83}  & \textbf{10.39}  & \textbf{8.75} & \textbf{9.84}  & \textbf{9.84}  & \textbf{8.54} & -0.03 \\
dprnn & FOMAML & \checkmark & -    & \textbf{7.9}  & \textbf{11.77} & \textbf{5.41} & \textbf{10.27} & \textbf{8.88} & \textbf{7.54} & \textbf{11.63} & \textbf{10.2} & \textbf{11.84} & \textbf{11.31} & \textbf{9.21}    & 0.03 \\
\midrule
\midrule
conv  & joint  & - & \checkmark & 4.73	& 7.63	& 2.35	& 6.02	& 5.63	& 2.92	& 5.03	& 5.74	& 7.55	& 7.11	& 5.38 & 0.18\\
conv  & joint  & \checkmark & \checkmark  & 4.72	& 7.23	& 2.48	& 6.14	& 5.15	& 2.82	& 5.48	& 5.56	& 7.09	& 7.17	& 5.28 & 0.22\\
dprnn & joint  & - & \checkmark & 5.43	& 8.52	& 3.78	& 7.46	& 6.46	& 4.71	& 7.10	& 6.96	& 8.54	& 8.88	& 6.61 & -0.10\\
dprnn & joint  & \checkmark & \checkmark  & 5.61	& 8.48	& 3.62	& 7.42	& 6.76	& 4.84	& 8.61	& 6.86	& 8.58	& 8.99	& 6.66 & -0.07\\
\midrule
conv  & FOMAML & \checkmark & \checkmark & 5.90	& 7.81	& 3.79	& 7.09	& 6.44	& 4.75	& 3.65	& 6.66	& 7.46	& 7.60	& 6.31 & 0.05\\
conv  & MAML   & \checkmark & \checkmark & \textbf{6.95}	& \textbf{10.3}	& \textbf{4.70} & \textbf{9.22} & \textbf{7.67}	& \textbf{6.98} & \textbf{11.08} & \textbf{8.8} & \textbf{10.34} & \textbf{9.87} & \textbf{8.08} & 0.02\\
dprnn & FOMAML & \checkmark & \checkmark & \textbf{7.31}	& \textbf{8.86}	& \textbf{5.13}	& \textbf{8.59}	& \textbf{8.05}	& \textbf{6.17}	& \textbf{9.16}	& \textbf{7.87}	& \textbf{8.90}	& \textbf{9.01}	& \textbf{7.71} & -0.02\\
\bottomrule
\end{tabular}
\caption{Results of joint training and MAML methods under the 3-speaker setting on the 10 target accents of the Speech Accent Archive. For the Conv-TasNet(conv) and Dual-path RNN model(dprnn), we mark the highest Si-SNRi values bold. The right most column is the Pearson correlation between accent similarities and performance.
}
\label{tab:accent_results_3spk}
\end{table*}
\begin{table}[h]
\centering
\setlength\tabcolsep{1.0pt}
\begin{tabular}{cccc|ccc|ccc}
\toprule
      &        &   &    & \multicolumn{3}{c|}{2-speaker}  & \multicolumn{3}{c}{3-speaker}  \\
model & method & f.t. & noise & zhCN & zhHK & zhTW & zhCN & zhHK & zhTW \\
\midrule
conv  & joint  & - & -  & 5.69     & 6.87     & 5.27     & 3.82     & 5.47     & 3.47     \\
conv   & joint  & \checkmark  & - & 5.93     & 6.78     & 5.38     & 4.05     & 5.42     & 3.61     \\
dprnn & joint  & - & -  & 6.71     & 7.47     & 5.75     & 4.49     & 6.45     & 4.66     \\
dprnn & joint  & \checkmark & -  & 6.7      & 7.36     & 5.72     & 4.65     & 6.6      & 4.79     \\
\midrule
conv  &  FOMAML & \checkmark & -  & 7.19     & 6.91     & 6.46     & 5.12     & 6.54     & 4.77     \\
conv  & MAML & \checkmark & -  & \textbf{7.91}     & \textbf{8.49}     & \textbf{7.45}     & \textbf{6.36}     & \textbf{7.66}     & \textbf{6.53}    \\
dprnn & FOMAML & \checkmark & -  & \textbf{7.89}     & \textbf{8.22}     & \textbf{6.48}     & \textbf{5.59}     & \textbf{7.77}     & \textbf{6.29}     \\
\midrule
\midrule
conv  & joint  & - & \checkmark   & 4.36	& 5.38     & 4.16      & 3.27 & 4.83  & 3.05     \\
conv   & joint  & \checkmark & \checkmark   & 4.60   & 5.40   & 4.18     & 3.43 & 4.75   & 3.15     \\
dprnn & joint  & - & \checkmark   & 5.24 & 5.91     & 4.72     & 3.90 & 5.67     & 4.10     \\
dprnn & joint  & \checkmark & \checkmark   & 5.22 & 5.62  &4.53      & 4.00   & 5.73     &4.11      \\
\midrule
conv  & FOMAML & \checkmark & \checkmark   &5.99 & 5.73     & 5.21     & 4.58     & 5.86     & 4.32 \\
conv  & MAML & \checkmark & \checkmark   & \textbf{6.29}    & \textbf{6.59}    & \textbf{5.34}     & \textbf{4.91}    & \textbf{6.81}     & \textbf{5.49}    \\
dprnn & FOMAML & \checkmark & \checkmark   & \textbf{6.56}   & \textbf{6.89}     & \textbf{6.21}      & \textbf{5.81}    & \textbf{6.98}  & \textbf{5.93}    \\
\bottomrule
\end{tabular}
\caption{Results of joint training and MAML methods on the testing split of the three CommonVoice Chinese speech data, zh-CN, zh-HK and zh-TW. For the Conv-TasNet(conv) and Dual-path RNN(dprnn) model, we mark the highest Si-SNRi values bold.}
\label{tab:CV_results}
\vspace{-4mm}
\end{table}

\subsection{MAML methods vs. Joint training}
Table \ref{tab:accent_result_2spk} shows the results of different training methods, joint training, FOMAML, and MAML under the 2-speaker setting. 
From the results of joint training, we found out that in some cases, fine-tuning the jointly trained models helps improve the performance, while some do not. 
For the clean speech of accents Ga and Ku, performances degrade after fine-tuning no matter which model we use. A similar phenomenon is observed under the 3-speaker setting in Table \ref{tab:accent_results_3spk} for accents Ga, Li and Ta. This implies that jointly trained speech separation models are not guaranteed better performance after adaptation. 
For MAML, we can see that in most cases, MAML methods outperform joint training methods. Even under noisy settings, MAML methods still perform better. 
The results suggest that the initialized parameters obtained from MAML can be adapted better to the new unseen accents and are more robust to noise compared with joint training. 

Table~\ref{tab:CV_results} is the results of evaluating on the CommonVoice Chinese speech dataset, zh-CN, zh-HK, and zh-TW. 
Different from the previous experiments, here we adapt models to languages unseen in training tasks. 
MAML methods also dominate joint training on the three Chinese accents (zh-CN, zh-HK, and zh-TW).
The results show that MAML can better adapt model parameters to both unseen accents and unseen languages.

For Tables \ref{tab:accent_result_2spk}-\ref{tab:CV_results}, MAML methods outperform joint training no matter we use the Conv-TasNet model or Dual-path RNN model, indicating that the MAML methods are indeed model agnostic.

\subsection{MAML vs. FOMAML}
From Tables \ref{tab:accent_results_3spk} and \ref{tab:CV_results}, training the Conv-TasNet model with MAML outperforms training with FOMAML on every accent. 
This shows that under most of the settings, MAML is a better method for multi-accent speech separation compared to its simplified version, FOMAML. 
From Table \ref{tab:accent_result_2spk}, for the Conv-TasNet model, training with MAML and FOMAML may have similar performance under the 2-speaker setting, but there is a vast difference between the computational time. 
We evaluated the computational time for a meta batch to finish one inner loop and outer loop step. 
Training with MAML(19021 ms) requires 15 times more computational time compared to FOMAML, so there is a trade off between the accuracy and computational time.



\subsection{Accent similarity and performance}
The rightmost column of Tables \ref{tab:accent_result_2spk} and \ref{tab:accent_results_3spk} show the Pearson correlations of the speech separation performance and the native-likeness of the utterances in the Speech Accent Archive. 
The correlation values of joint training are mostly within $[-0.4, 0.4]$, which shows that there is correlation between the two variables, but it is weak and likely to be viewed as unimportant. 
This suggests that the Si-SNRi values do not relate much to the native-likeness of accents, indicating that the similarity between the testing accents and the training data does not largely affect the performance in this speech separation task.
Furthermore, MAML methods have a correlation approaching zero, while joint training methods still relate to the accent similarities to some extent.
This implies that MAML methods are more accent agnostic compared to joint training. 
These results are consistent even when noise is involved.

\section{Conclusion}
\label{conclusion}
In this paper, we trained SS models with only native English speech. We conclude that MAML methods outperform joint training on adapting speech separation models with unseen accents, languages and noise. We found that the transferability to new accents during the testing phase is not dominated by the similarity of accents between the source data and target data. In the future, we will modify the setting of the meta task construction in the hope of improving the SS task.  


\bibliographystyle{IEEEbib}
\bibliography{strings,refs}

\end{document}